\documentclass[sigconf]{acmart}

\usepackage{ulem}
\usepackage{amsmath}
\usepackage{multirow}

\AtBeginDocument{%
  \providecommand\BibTeX{{%
    \normalfont B\kern-0.5em{\scshape i\kern-0.25em b}\kern-0.8em\TeX}}}

\copyrightyear{2024}
\acmYear{2024}
\setcopyright{acmlicensed}\acmConference[WWW '24 Companion]{Companion Proceedings of the ACM Web Conference 2024}{May 13--17, 2024}{Singapore, Singapore}
\acmBooktitle{Companion Proceedings of the ACM Web Conference 2024 (WWW '24 Companion), May 13--17, 2024, Singapore, Singapore}
\acmDOI{10.1145/3589335.3648301}
\acmISBN{979-8-4007-0172-6/24/05}

\begin{document}

\title{Scaling User Modeling: Large-scale Online User Representations for Ads Personalization in Meta}





\author{Wei Zhang}
\authornote{Both authors contributed equally to this paper.}
\email{weizhng@meta.com}
\author{Dai Li}
\email{daili1@meta.com}
\authornotemark[1]
\orcid{0009-0000-6645-5620}
\author{Chen Liang}
\author{Fang Zhou}
\author{Zhongke Zhang}
\author{Xuewei Wang}
\affiliation{%
  \institution{Meta Platforms, Inc.}
  \city{Menlo Park}
  \state{CA}
  \country{USA}
}

\author{Ru Li}
\author{Yi Zhou}
\author{Yaning Huang}
\author{Dong Liang}
\author{Kai Wang}
\author{Zhangyuan Wang}
\author{Zhengxing Chen}
\affiliation{%
  \institution{Meta Platforms, Inc.}
  \city{Menlo Park}
  \state{CA}
  \country{USA}
}

\author{Fenggang Wu}
\author{Minghai Chen}
\author{Huayu Li}
\author{Yunnan Wu}
\author{Zhan Shu}
\author{Mindi Yuan}
\author{Sri Reddy}
\email{sriharir@meta.com}
\affiliation{%
  \institution{Meta Platforms, Inc.}
  \city{Menlo Park}
  \state{CA}
  \country{USA}
}

\renewcommand{\shortauthors}{Wei Zhang et al.}

\begin{abstract}
Effective user representations are pivotal in personalized advertising. However, stringent constraints on training throughput, serving latency, and memory, often limit the complexity and input feature set of online ads ranking models. This challenge is magnified in extensive systems like Meta’s, which encompass hundreds of models with diverse specifications, rendering the tailoring of user representation learning for each model impractical. To address these challenges, we present Scaling User Modeling (SUM), a framework widely deployed in Meta’s ads ranking system,  designed to facilitate efficient and scalable sharing of online user representation across hundreds of ads models. SUM leverages a few designated upstream user models to synthesize user embeddings from massive amounts of user features with advanced  modeling techniques. These embeddings then serve as inputs to downstream online ads ranking models, promoting efficient representation sharing. To adapt to the dynamic nature of user features and ensure embedding freshness, we designed SUM Online Asynchronous Platform (SOAP), a latency-free online serving system complemented with model freshness and embedding stabilization, which enables frequent user model updates and online inference of user embeddings upon each user request. We share our hands-on deployment experiences for the SUM framework and validate its superiority through comprehensive experiments. To date, SUM has been launched to hundreds of ads ranking models in Meta, processing hundreds of billions of user requests daily, yielding significant online metric gains and improved infrastructure efficiency. 

\end{abstract}





\maketitle

\section{Introduction}
Personalization~\cite{personalization1, personalization2, personalization3, collaborativefilter, pinnerformer, ali}  is the cornerstone of modern online advertising, enhancing both advertiser returns and user experiences.
At the heart of personalization is user understanding, which traditionally relied on manually engineered features and simplistic architectures.
The advent of deep learning-based recommender systems has shifted this paradigm, leveraging sophisticated neural network models to learn intricate user representations ~\cite{deep_wide, fm1, fm2, collaborativefilter, dcn, dcnv2, deepFM, airbnb, tencent, pinnerformer, ali, pinsage, twhin, pinterest_transact, yuan1, yuan2}. 
However, practical constraints such as training throughput, serving latency, and host memory limit their capacity to fully utilize the plethora of user data. In expansive systems like Meta's, which encompass lots of diverse models processing hundreds of billions of user requests daily, these limitations become even more pronounced, leading to several interrelated challenges in effective user representation learning:

\begin{figure}
  \centering
  \includegraphics[width=\linewidth]{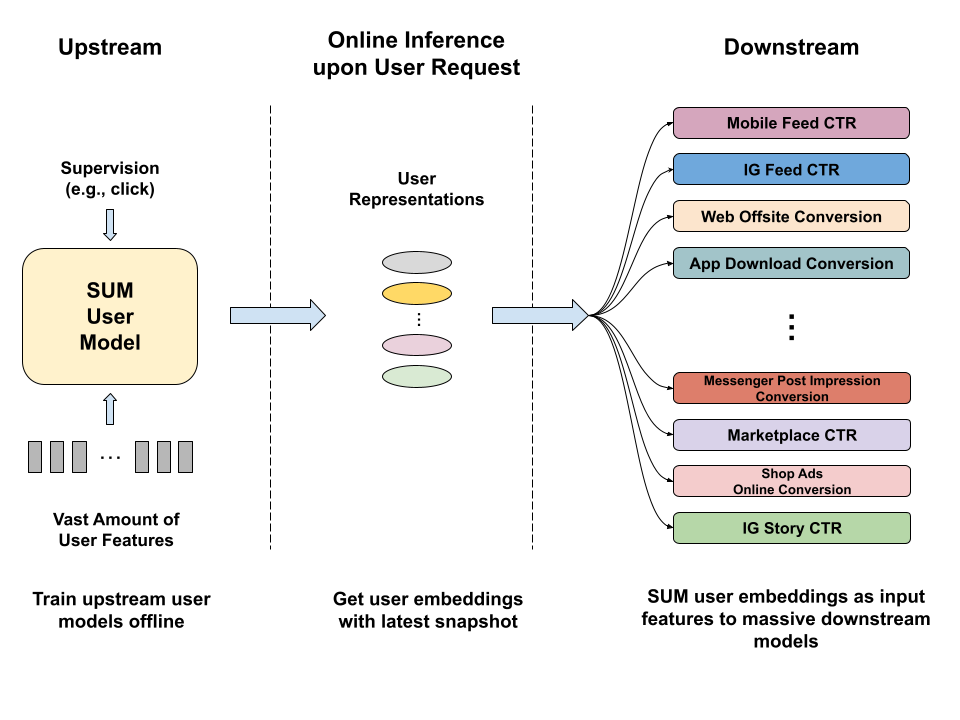}
  \caption{An overview of the proposed SUM framework. SUM envisions the following state: We have a few dedicated user models that can consume a vast amount of user-side features with advanced user modeling techniques and produce embedding representation for each user. User models can be trained with multiple supervisions (click, conversion, etc.) and support recurring snapshot updates. Multiple downstream models are able to safely consume user model output (i.e., SUM user embeddings) as input features. As a result, the gain from the user model will add up across all the downstream models.}
  \label{fig:concept}
\end{figure}

\begin{itemize}
    \item \textbf{Sub-optimal Representations}: Models independently learning user representations often yield inferior results.
    \item \textbf{Feature Redundancy}: Overlapping user features across models necessitate unnecessary duplication in training pipelines, leading to high storage overhead. 
    \item \textbf{Data Scarcity for Specialized Models}: Models serving niche segments lack the large training data volumn needed for robust user understanding.
    \item \textbf{Intensive Tailoring}: Customizing architecture and feature selection for specific performance needs of each model is not scalable.
\end{itemize}
Addressing these challenges is crucial for supporting user modeling with the growing complexity and efficiently scaling representation learning advances across numerous models. 

To address these challenges, we introduce \textit{Scaling User Modeling (SUM)}, an online framework revolutionizing user modeling within Meta Ads. SUM is designed to capitalize on advanced modeling techniques while adhering to practical constraints and promoting efficient, scalable representation sharing across models.

SUM employs an upstream-downstream paradigm (Figure~\ref{fig:concept}), drawing inspiration from recent works in user history modeling ~\cite{ali, tencent, pinnerformer}. Our approach involves training a small number of large-scale upstream user models with sophisticated architectures under diverse supervisions, like clicks and conversions. The user models process a vast amount of user-side signal (features) to synthesize compact user embeddings (representations). These embeddings are then seamlessly integrated into various downstream production models, propagating advanced user modeling and representation sharing.

A key design in SUM is its adaptability to the dynamic nature of user features. Our serving system, SUM Online Asynchronous Platform (SOAP), is a pivotal component, enabling latency-free, asynchronous serving that achieves embedding freshness while overcoming the latency limit regardless of model complexity, substantially surpassing traditional offline-based solutions. Complementing SOAP, we've devised a recurrent training regimen, enhanced by average pooling techniques, to maintain model freshness and stabilize user embeddings.

Since its initial deployment, SUM has been launched into hundreds of production ads ranking models in Meta, leading to notable improvements in both offline and online business metrics. Its efficacy and adaptability have also prompted its extension into Meta's feed and quality models. The primary contributions of this paper are:

\begin{itemize}
    \item We propose SUM, an online user modeling framework that focuses on optimal usage of all existing user features and complex models within production constraints, and scaling representation learning across hundreds of models.  
    \item We detail a large-scale user model architecture, essential for learning meaningful user representations from a vast array of input user features, which serves as the backbone of SUM models. 
    \item We present SOAP, a design of latency-free online serving system that balances feature freshness against serving constraints, surpassing offline solutions and enhancing scalability.
    \item We provide extensive experimental studies and practical insights from deploying SUM, including strategies like average pooling to mitigate embedding distribution shifts from frequent model updates. 
\end{itemize}

\section{Related Work}
\begin{figure*}
  \centering
  \includegraphics[width=\linewidth]{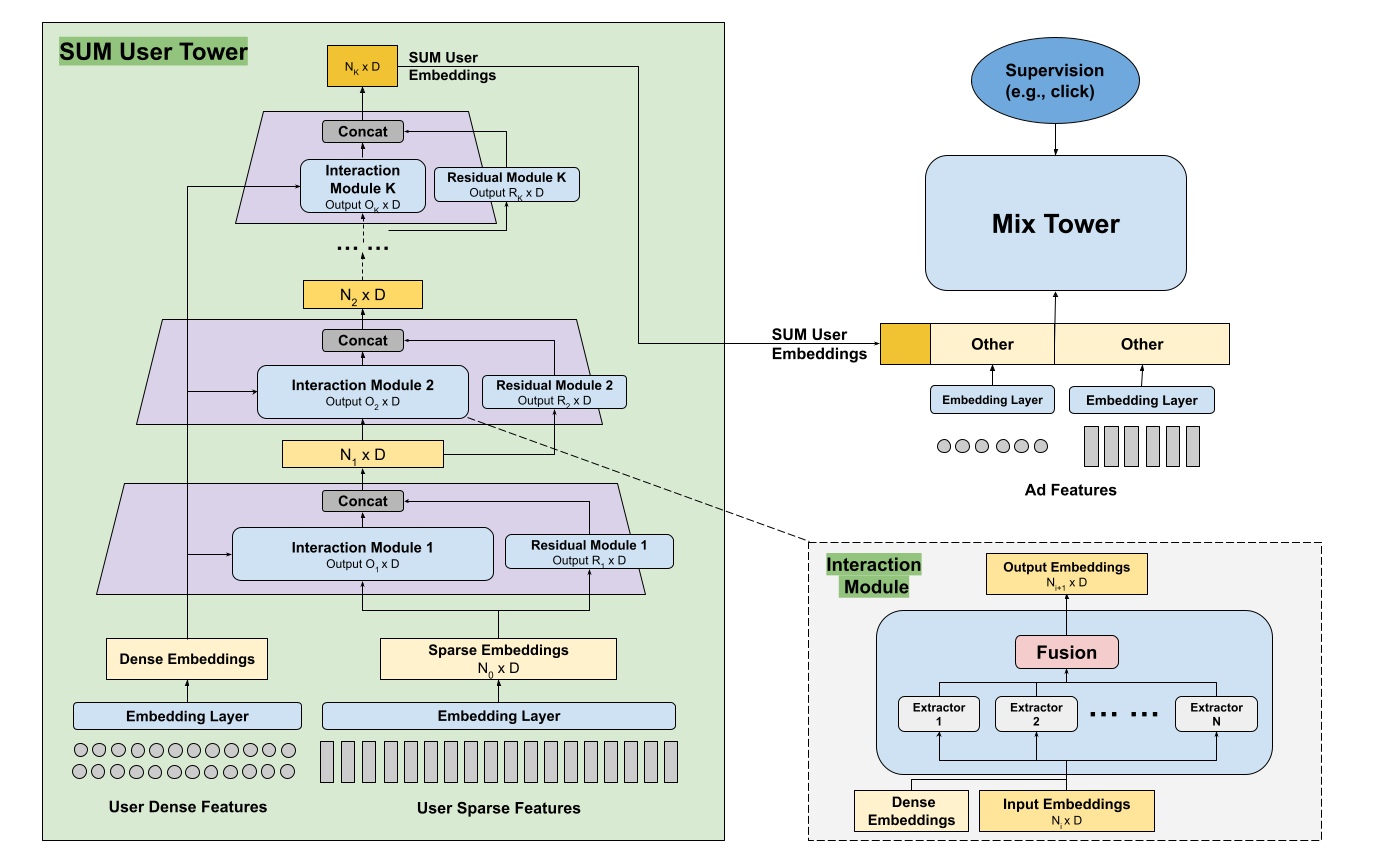}
  \caption{An illustration of SUM upstream model architecture. The SUM user tower consumes the massive amount of user features and outputs a few user embeddings which will then be fed to mix tower. The user tower is the core of the upstream model and has a pyramid architecture with residual connections to learn user representations gradually. Its basic building block, Interaction Module, consists of various feature extractors in parallel to capture different feature interactions.}
  \Description{A woman and a girl in white dresses sit in an open car.}
  \label{fig:user_tower}
\end{figure*}
Personalization has been at the forefront of ads ranking and recommender system research. A plethora of modeling techniques have been proposed to deliver customized ads for users. Factorization Machines~\cite{fm1, fm2} transform the high-dimensional and sparse features into low-dimensional vectors using matrix factorization. Collaborative Filtering~\cite{cf1, cf2, cf3} leverages the patterns of similar users and items to predict personalized recommendations for users. The proliferation of deep learning has empowered modern ads ranking models and recommender systems to learn high-order and non-linear interactions from large-scale datasets~\cite{deep_wide, deepFM, dcn, dcnv2, neural_cf}. Cutting-edge models like Recurrent Neural Networks (RNNs)~\cite{rnn}, Transformers~\cite{pinterest_transact, pinnerformer} and Graph Neural Networks (GNNs)~\cite{twhin, aliitemgraph} have further advanced the depth and breadth of personalization.

Nevertheless, stringent infrastructural constraints often limit the sophistication of model architectures and the spectrum of user features in online models, inhibiting optimal user representation. As a result, the industry has embraced embedding-based approaches that utilize an upstream-downstream paradigm~\cite{twhin, pinsage, itemsage, aliitemgraph, airbnb, pinnerformer, tencent, ali, pinterest_transact, yuan1, yuan2}. The upstream model is usually trained and updated offline to produce condensed embeddings, which will be used as input features for downstream online production models to consume. In this way, the latency constraints are relaxed, allowing for larger model complexity. The embeddings mainly contain item embeddings ~\cite{twhin, pinsage, itemsage, aliitemgraph, airbnb} and user embeddings ~\cite{pinnerformer, tencent, ali, pinterest_transact}. 

For item embeddings, TwHin from Twitter~\cite{twhin} derives embeddings from a heterogeneous knowledge graph, capturing nuances of users, tweets, and ads. PinSage from PinInterest ~\cite{pinsage} combines random walks with graph convolutions to learn graph node embeddings. Airbnb~\cite{airbnb} extracts listing embeddings to improve search ranking which are pre-computed offline and stored daily to make them real-time available during online serving. ItemSage~\cite{itemsage} leverages a transformer-based architecture to learn product embeddings from multi-modalities, which is computationally expensive. It adopts an offline serving solution that performs daily batch inference. Given the inherent stability of item features, embedding staleness from such offline serving solutions remains acceptable. But for the user features used by SUM, it hurts the user embedding performance significantly as discussed in later sections, which makes ensuring embedding freshness crucial for SUM productionization.

For user embeddings, PinnerFormer from Pinterest~\cite{pinnerformer} models user sequences, leaning on a daily offline batch setting to reduce infra pressure. Tencent~\cite{tencent} learns user app usage embeddings and updates them daily for the active users, while the upstream model updates are much less frequent to avoid embedding distribution shift and corresponding downstream model updates. Ali~\cite{ali} mines long-term user behavior through e2e learning, which decouples the user modeling part from entire model and stores the latest user embeddings for easy access online. The user embedding updates are triggered by user behavior events instead of traffic request. Previous studies mainly concentrated on learning user history behavior, along with additional efforts to build and maintain data pipelines for user-centric behaviors. While this work focuses on exploring a feasible solution for supporting user modeling with the growing complexity impractical to be directly added in production ads models, and efficiently scaling user representation sharing in large-scale ads ranking systems. 


\section{Model Architecture}

In this section, we present the design of the upstream SUM user model.
\subsection{Preliminary}
As depicted in Figure~\ref{fig:user_tower}, the SUM user model utilizes the acclaimed DLRM architecture~\cite{dlrm}. It is divided into two main components: the user tower and the mixed tower. The user tower processes an extensive range of user-side input features, converting them into a refined set of SUM user embeddings. These embeddings are then relayed to the mixed tower, where they engage in further interactions with other types of features, such as those related to advertisements.

We classify input user features into two major categories: dense and sparse. Dense features are numerical or continuous variables, for example, the frequency of clicks on a particular page. Sparse features are typically categorical or binary variables with high cardinality, including user ID and page ID. These sparse features are storage and computation-intensive, necessitating embedding lookup tables with an exceedingly large parameter count to map raw values to dense embeddings~\cite{autoshard, he2014practical, deep_wide, youtube, DIN}. Standard models often handle fewer than 300 user sparse features, while more compact models are limited to under 100. In contrast, SUM user models are significantly more expansive, accommodating much more user features, far exceeding those managed by downstream models.

\subsection{User Tower}
The user tower employs a pyramid-like structure with successive Interaction Modules to compress these input features methodically. Residual connections~\cite{resnet} are applied to facilitate the training as well as retain the original information. After dense features and sparse features pass the initial embedding layers, assuming we have $N_0$ sparse embeddings and $N_{dense}$ dense embeddings of dimension $D$, the user tower coalesces them into $N_K$ output user embeddings of the same dimension, where $N_0 >> N_K$. For the n-th Interaction Module, 
\begin{equation}
    X_{n} = Concat(Interaction(X_{n-1}), X_{dense}), Residual(X_{n-1}))
\end{equation}
Interaction, Residual denote Interaction Module and Residual Module in Figure \ref{fig:user_tower} respectively. $X_{n-1}$ is the input of the n-th Interaction Module. $X_{dense}$$\in$$R^{N_{dense}\times D}$ is the raw dense embeddings after the initial embedding layer. Residual Module is usually MLPs. The Interaction Module integrates multiple parallel feature extractors, designed to capture a spectrum of complementary interactions. The feature extractor has different options, some are listed below.

\paragraph{MLP}
It's used to learn general nonlinear low-level implicit representations.

\paragraph{Dot Compression with Attention}
Compressing the pairwise dot product matrix is an effective way to increase both model capacity and efficiency. We incorporate an advanced dot compression with attention and residual connections to learn high-level explicit representations, formulated as 
\begin{equation}
  Y = MLP(Concat(LC(X_{dense}), LC(X)))
\end{equation}
\begin{equation}
    Z = MLP'(Concat(LC'(X_{dense}), LC'(X)))
\end{equation}
\begin{equation}
  output = X(X^TY + Z)
\end{equation}
where $FC(*)$ means a fully connected layer without non-linear function, $LC(*)$ means linear compression. $LC(X_{dense})=FC(X_{dense})$ for dense features. $LC(X)$ are a set of weighted sum's of the sparse embeddings in $X$. The reason we use $LC(*)$ before $Concat(*)$ is to reduce the input size for $MLP(*)$ to improve model efficiency. $Y$ is the attention weights. $Z$ is the residual branch.

\paragraph{Deep Cross Net}
DCN~\cite{dcnv2, dcn} learns expressive representations through effective explicit and implicit feature crosses. Its basic component is the cross layer which can be illustrated by the following eqation.
\begin{equation}
  X_{n+1}=X_{0} \odot\left(W_{n} X_{n}+\mathbf{b}_{n}\right)+X_{n}
\end{equation}
Here $W_n$ and $b_n$ are learnable weight and bias. By stacking multiple cross layers, the model can capture high-order verctor-level and bit-level interactions.

\paragraph{MLP-Mixer}
MLP-Mixer~\cite{mlpmixer} is an all-MLP architecture that was originally designed for Computer Vision. This architecture can be viewed as a unique CNN, which uses 1x1 convolutions for channel mixing, and single-channel depth-wise convolutions for token mixing as shown in Equation \ref{eq:channel} and Equation \ref{eq:token}, respectively.
\begin{equation}
    Y = X + W_2 \cdot \text{ReLU}\left(W_1 \cdot \text{LayerNorm}(X)^T\right)^T
\label{eq:token}
\end{equation}
\begin{equation}
    Z = Y + W_4 \cdot \text{ReLU}\left(W_3 \cdot \text{LayerNorm}(Y)\right)
\label{eq:channel}
\end{equation}
$W_1$, $W_2$, $W_3$, $W_4$ are learnable weights.

\subsection{Mix Tower}
The mix tower adopts a DHEN-style architecture~\cite{dhen}. It does not contain any user-side features as input apart from the SUM user embeddings from user tower. This strategic design choice aims to bolster upstream training throughput and encourage the model to refine user representations primarily through the user tower. Our experience indicates that a more intricate mix tower architecture, though shows stronger predictive performance of the upstream model, does not bring notable improvements to the quality of user embeddings or yield downstream benefits. We utilize a multi-task cross-entropy loss:
\begin{equation}
    \mathcal{L} = -\frac{1}{N} \sum_{i=1}^{N} \sum_{t=1}^{T} w_t \left( y_{ti} \log(\hat{y}_{ti}) + (1-y_{ti}) \log(1 - \hat{y}_{ti}) \right) 
\end{equation}
where $w_t$ is the weight for task $t, t = 1, 2, ... T$, representing its importance in the final loss. $y_{ti}\in\{0, 1\}$ is the label for sample $i$ in task $t$. $\hat{y}_{ti}$ is the predicted value of the model for sample $i$ in task $t$. $N$ is the number of samples.

\section{Online Serving System: SOAP}
In contrast to the relatively stable item features used by~\cite{itemsage, twhin, pinsage}, the user features within the SUM user tower often changes drastically. A considerable portion of these features are categorical variables, which face challenges from both the introduction of new ids and shifting semantic meanings of existing ids. In this case, the embedding staleness from offline-based serving solutions, either batch inference~\cite{itemsage, pinnerformer} or event-triggered inference~\cite{ali} can detrimentally impact downstream performance as discussed in Section 6.4. Thus we decide to implement online inference of the user embeddings upon each user request. However, the latency budget is usually tight for online inference. Typically, once a user request is received, there is a mere 30ms window to conduct the inference and relay user embeddings to downstream ranking models, thereby limiting the complexity of the user model and, by extension, the representative power of SUM user embeddings. 

\begin{figure}[h]
  \centering
  \includegraphics[width=\linewidth]{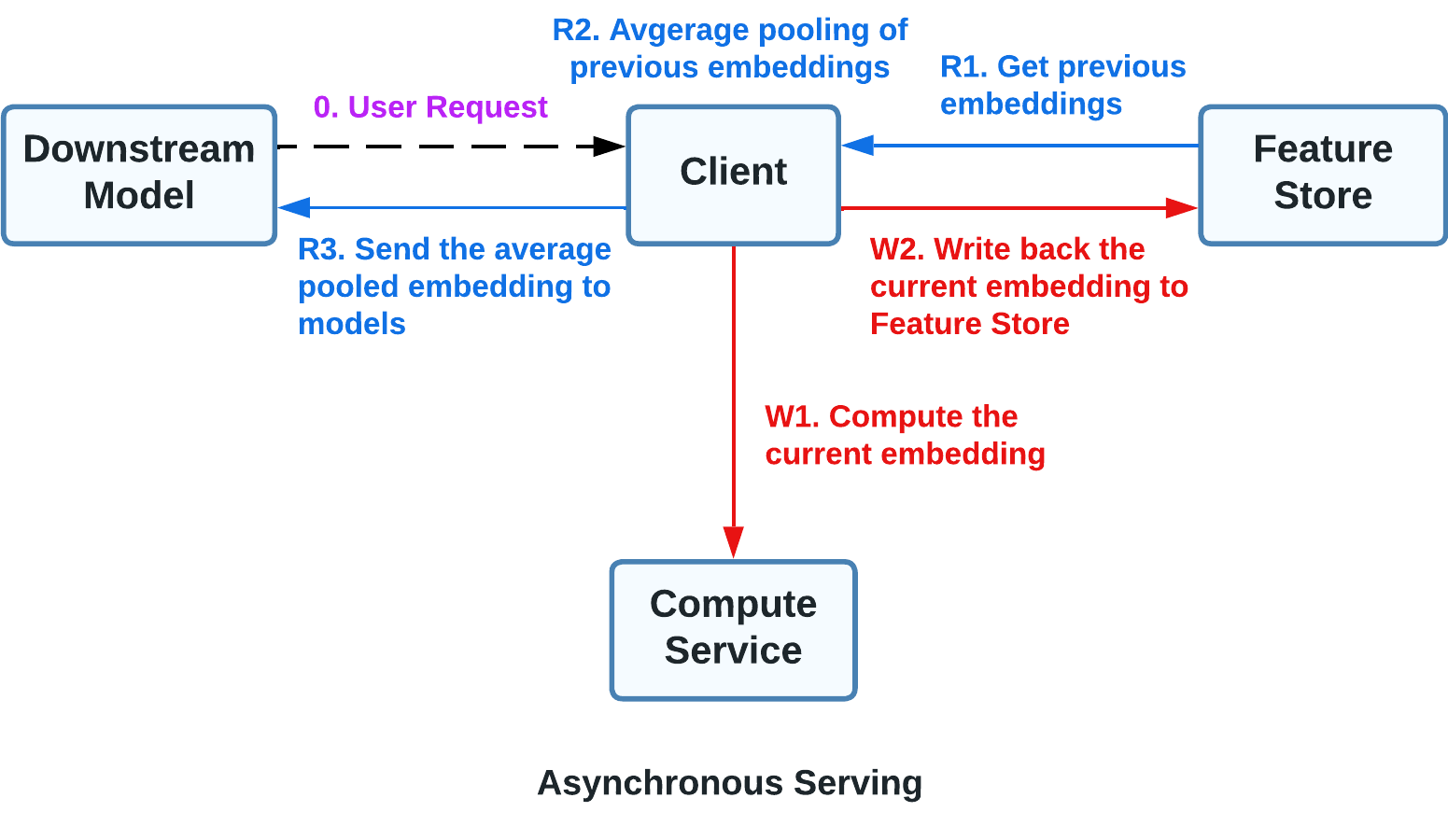}
  \caption{An illustration of SOAP, the online serving system for SUM, which leverages our proposed Async Serving paradigm.}
  \Description{A woman and a girl in white dresses sit in an open car.}
  \label{fig:serving}
\end{figure}
To tackle the aforementioned challenges, we designed SUM Online Asynchronous Platform (SOAP) which harnesses a novel Async Serving paradigm tailored for SUM online inference. As depicted in Figure~\ref{fig:serving}, upon the downstream model receiving a user request, the SOAP Compute Center computes the current embedding using the latest snapshot, subsequently writing it to the Feature Store. In parallel, the Client immediately reads the previous embeddings for this user from Feature Store and forwards the average pooled embedding to the downstream model, without waiting for the Compute Center's inference to conclude. In this way, SOAP decouples the feature write path from the read path. The calculation and updating of the current embedding is invoked as an asynchronous call. By separating the expensive and latency-intensive write path from the read path, Async Serving effectively removes the latency limit and can in theory support complex user models with arbitrarily long inference latency (in practice, the inference latency will still proportionally affect the serving power use, but it is no longer a hard blocker). It significantly outperforms offline-based methods, while trades off a slight amount of realtime-ness to boost scaling potential yeilding improved ROI, as elaborated in Section 6.4.

To further reduce inference time, only the user tower of the SUM user model is served in SOAP to generate user embeddings, instead of the entire model. 

\section{Productionization}
\subsection{Model Training}
The SUM user model is trained offline in a recurring manner which allows the model to retain historical patterns while incrementally adapting to envolving user preferences. This frequent snapshot publishing, combined with online inference, empowers SUM to consistently deliver up-to-date user embeddings to downstream models.
\subsection{Embedding Distribution Shift}
However, with recurring training, the user model keeps publishing new snapshots. This implies that even for the same inference input, the computed user embeddings are changing over time. We call this phenomenon "embedding distribution shift". As shown in Section 6.5, it significantly hurts the effectiveness of the user embeddings. There are two options to mitigate embedding distribution shift. 

\textbf{Option 1} One is to make sure a new version of embedding features has been seen by downstream models during training, before it is used in inference. This option requires changes in model training, serving and feature logging, which brings additional complexity to our system. 

\textbf{Option 2} The other option is to reduce distribution shift by enforcing a new version to be similar to its previous version. This can be done through various ways such as regularization ~\cite{twhin}, distillation and post processing with average pooling. 

We propose to mitigate this by average pooling over the 2 most recent cached embeddings and the current computed embedding as the final current embedding of that user. Average pooling has been a widely adopted practice in ads ranking and recommender systems, especially to cope with code-start problems ~\cite{airbnb, average1, average2}. We choose this option because it is cheap, brings good performance, and increases feature coverage for free.
\subsection{Feature Storage Optimization}
In production, we usually configure K=2 and D=96, meaning a user model generates 2 SUM user embeddings, each of dimension 96. After investigating the value of K, we found K=2 has good ROI between performance enhancement and feature storage efficiency. To further reduce the online and offline storage use, the two user embeddings are quantized from fp32 to fp16; we evaluated that this quantization didn't lead to downstream performance difference. 
\subsection{Distributed Inference}
To lift memory constraints of SUM user model for more performance gains, we've incorporated Distributed Inference (DI) for the user tower so that the online inference workload could be efficiently distributed among multiple hosts.

\section{Experiments}
\subsection{Experimental Setup}
\subsubsection{Dataset} In this work, all the experiments were done on industrial datasets. We did not use public datasets as there are gaps applying them onto our serving system and the large discrepancy with internal models makes them unsuitable for downstream experiments.  
\subsubsection{Evaluation Metric}

\paragraph{Normalized Entropy} We use Normalized Entropy (NE)~\cite{he2014practical,zhang2022dhen} defined in \eqref{eq:ne} as the metric to evaluate model predictive performance offline. It measures how accurately a model is predicting when users will click on ads. It is equivalent to the average logarithmic loss per impression divided by what the average logarithmic loss per impression would be if a model predicted the background Click Through Rate (CTR, i.e., a constant model that always predicts the average CTR) for every impression. Lower is better. 
\begin{equation}
NE = \frac{\frac{-1}{N} \sum_{i=1}^{n} \left( y_i \log(p_i) + (1-y_i) \log(1 - p_i) \right)} {- (p \log(p) + (1 - p) \log(1 - p))}
\label{eq:ne}
\end{equation}
Here $N$ represents the total number of examples in the dataset. $y_i\in\{0, 1\}$ are the labels, $i = 1, 2, ..., N$. $p_i$ is the estimated probability of a click for each impression, while p is the average empirical CTR. \\
\paragraph{Feature Importance Ranking} We use the Feature Importance Ranking (FI) ~\cite{anaraki2019feature} to evaluate how important one feature is to a model. SUM embeddings generated by the user model are fed to various production models as input. By analyzing the FI ranks of SUM embeddings compared to other available features of one model, we can understand the relative value and impact of SUM user embeddings. 
\subsubsection{FB CTR SUM User Model} In practice, we maintain a few SUM user models, each trained on different datasets and supervisions. As a representative example for our ensuing discussions, we will focus on the FB CTR SUM user model. FB CTR SUM user model utilizes the training dataset of Facebook mobile app feed, comprising roughly 6 billion daily examples, and targets the CTR prediction task. The user tower consists of four sequentially stacked Interaction Modules, employing MLP, Dot Compression with Attention, and the MLP-Mixer as feature extractors. It processes around 600 user-side sparse features and 1,000 user-side dense features. As a result, the user tower occupies a size of 160 GB and requires 390M inference FLOPs.

\subsection{Downstream Offline Results}
\begin{table}
  \begin{tabular}{p{1.1in}ccl}
    \toprule
    Task & Domain & NE diff (\%) & FI \\
    \midrule
    CTR&Mobile feed & -0.19 & 2\\
    CTR&Home feed & -0.12 & 2\\
    CTR&Mobile story & -0.16 & 1\\
    CTR&Dynamic ads & -0.12 & 4\\
    CTR&Marketplace & -0.08 & 3\\
    CTR&Channel watch & -0.12 & 2\\
    CTR&Instagram story & -0.05 & 15\\
    CTR&Messenger inbox & -0.35 & 1\\
    Post-imp offsite CVR & Mobile feed & -0.09  & 16\\
    Post-click offsite CVR& Mobile feed & -0.05 & 13\\
    MAI CVR & Mobile feed & -0.04 & 24\\
    Inline CVR & Mobile feed & -0.09 & 5\\
    \bottomrule
  \end{tabular}
  \caption{The performance of FB CTR SUM embeddings on some production ads ranking models in Meta. Offsite, MAI (mobile app install), and inline are all particular ad conversion types. CVR means conversion rate. NE diff means the NE difference of the model with FB CTR SUM embeddings compared to the model without them. Lower is better. FI is the Feature Importance ranking of SUM embeddings in all available sparse and embedding features of that model. The total number of ranked features here is usually thousands. Note that the model in production can only use a small portion of them due to aforementioned practical constraints.}
  \label{tab:overallimpact}
\end{table}

Table~\ref{tab:overallimpact} lists the offline NE performance of FB CTR SUM on some downstream models. From Table~\ref{tab:overallimpact}, we can see that SUM embeddings bring statistically significant NE gains across diverse downstream tasks such as CTR prediction, inline conversion prediction, mobile app install conversion prediction, and offsite conversion prediction in various domains. 

This underscores the strong representative power and generalizability of SUM. It is expected that they show higher gains in CTR tasks since they are trained on CTR data. One special case is the relatively smaller gains on Instagram models, highlighting the need for a Instagram SUM model to bridge the domain discrepancy between FB and Instagram. Additionally, on those small models like Messenger inbox with low model complexity and feature count, the benefits from large-scale user representation sharing are more salient. What is worth noticing is that, adding SUM embeddings brings minimal training throughput or other infra metrics change to the downstream models, given embedding features are dense representations that do not require computationally intensive embedding look-up tables necessary for sparse features. 

\subsection{Online Performance}

SUM has been launched to hundreds of production ads models in Meta, achieving significant gains in both offline and online business metrics, benefiting diverse platforms such as IG, FAM, Shop Ads, Messenger Ads, to name a few. 
Online A/B tests show that SUM led to 2.67\% online ads metric gains in total (0.2\% gain can be considered as statistically significant internally). 
Notably, while accomplishing these gains, SUM successfully avoided 15.3\% serving capacity increase, compared to directly introducing the same complexity into each downstream model. 

\subsection{Async Serving}

\begin{table}
  \begin{tabular}{cc}
    \toprule
    Model setting & NE diff(\%) \\
    \midrule
    Baseline, no SUM & 0\\
    Baseline + SUM (Frozen, 1-month staleness) & -0.034\\
    Baseline + SUM (Offline Batch, 1-day staleness) & -0.094\\
    Baseline + SUM (Online Realtime Serving) & -0.141\\
    Baseline + SUM (Online Async Serving) & -0.126\\
    \bottomrule
\end{tabular}
\caption{Comparison of SUM performances under different serving schemes. The staleness here is the number of days between the last training date of the user model and the first training date of the downstream ranking model.}
\label{tab:async}
\end{table}

We conduct experiments to compare 4 different serving solutions.\\
\textbf{Frozen User Model} The user model is only trained once and we keep using the initial snapshot to evaluate the new data and generate user embeddings. One important prerequisite for this setting to work is to only consume stable content-based features as input rather than short-lived ID features, which restricts the input feature space considerably. Given that our current user model setup tries to utilize all existing user-side information without differentiating stable features from non-stable features, the frozen model setting might not reach the full potential of the user models.\\
\textbf{Offline Batch} The user model is in daily recurring training after the initial training finished. We use the snapshot trained on (ds-1) to evaluate the data on (ds) to produce user embeddings. Typically there is usually 1 to 3 days staleness, contingent on the data pipeline design of the downstream models.\\
\textbf{Online Real-time Serving} Upon receiving a user request, the Client gets the current embedding from the Compute Service and the previous embeddings from Feature Store, then forwards the average pooled embedding as final current embedding to the downstream model. If the inference of Compute Service does not conclude within the stipulated latency window, it is considered a fallback.\\
\textbf{Online Async Serving} Details are described in Section 4. \\

For a fair comparison of real-time serving and Async serving, the user model we used for this set of experiments is a more compact version with only 20M inference FLOPs, as larger ones would have very high fallback rates. The results are presented in Table \ref{tab:async}, showing that transitioning the model serving from real-time to Async incurs a mere 10\% average loss, which is much smaller than the loss observed when shifting from real-time to the offline batch setting. Moreover, it paves the way for harnessing more intricate user models, underscoring the benefits of Async Serving.

\subsection{Embedding Distribution Shift}
We conduct offline experiments to understand the embedding distribution shift issue. Here the SUM user model is in Offline Batch mode for easier e2e experiments, i.e., we do daily recurring training for the user model and dump user embeddings using the daily updated moving snapshots. The two user embeddings are denoted as $Embedding 0$ and $Embedding 1$. For each user, we compare the cosine similarity and L2 norm change between the embeddings on consecutive dates, and report the average in Table~\ref{tab:cosineL2}. A side note is that, we also found that $Embedding 1$ brings larger training NE gain to the downstream model than $Embedding 0$, but $Embedding 0$ can bring additional gains on top of $Embedding 1$. Understanding why one embedding is much more stable than the other and why the more stable one brings larger training gains is not the focus of this work. It can be a potential area for future improvement.

\begin{table}
  \begin{tabular}{ccp{0.5in}p{0.5in}}
    \toprule
    Embedding & Avg pooling & Cosine & L2 norm \\
    \midrule
    \multirow{2}{*}{Embedding 0} & None & 0.87 & 13\% \\
    ~ & 3 & 0.97 & 5\% \\
    \hline
    \multirow{2}{*}{Embedding 1} & None & 0.97 & 5\% \\
    ~ & 3 & 0.99 & 2\% \\
  \bottomrule
  \end{tabular}
  \caption{Embedding distribution shift study. Embedding 0 and 1 are the two user embeddings output by a SUM upstream model. The number in Average pooling column means how many versions of the embedding are pooled to get final embedding. None means no average pooling applied. (Average) cosine similarity and (average) L2 norm change measure the similarity between two user embeddings produced by snapshots from consecutive dates, which can be regarded as the degree of embedding shift.}
  \label{tab:cosineL2}
\end{table}

As shown in Table~\ref{tab:es}, without average pooling, good training NE gain is observed, which means during training the downstream model can adapt to embedding distribution shift. However, evaluation NE gain is much smaller, which is expected: For evaluation, the embedding features are generated by a new user model snapshot, and the new version of embeddings is not seen by the downstream model during training. The sudden change leads to worse eval NE. Average pooling over 3 embeddings can significantly reduce embedding shift and improve performance, especially in terms of eval NE. 
\begin{table}
  \begin{tabular}{p{1.2in}p{0.7in}p{0.7in}}
    \toprule
    Model setting & Train NE (\%) & Eval NE (\%)\\
    \midrule
    Baseline & 0 & 0\\
    Baseline + SUM w/o AP & -0.081 & -0.014\\
    Baseline + SUM w/ AP & \textbf{-0.109} & \textbf{-0.127}\\
  \bottomrule
  \end{tabular}
  \caption{Downstream model performance of average pooling. AP stands for average pooling. The NE numbers here are NE difference compared to the baseline.}
  \label{tab:es}
\end{table}

\section{Conclusion}
In this paper, we present SUM, a large-scale online personalization framework which empowers Meta's ads ranking models with powerful user representations synthesized from a few SUM user models. Our innovative SOAP serving system facilitates online Async inference of SUM user embeddings, complemented with user model freshness and embedding stabilization. This not only enhances embedding freshness compared to offline serving methods but also unlocks the potential of employing more intricate user models. We delve into the challenges and our best practices in deploying SUM. The experimental results, coupled with its successful launches within Meta, attest to SUM's superiority. 

SUM has been launched to hundreds of production ads models in Meta, processing hundreds of billions of user requests daily. We believe that this work provides a practical, scalable and efficient path to improve ads personalization with effective user embedding sharing and address the inherent limitations of online ads models in terms of training throughput, memory and serving power. Looking ahead, we aspire to further improve the user modeling algorithm and the serving system to achieve even more nuanced user representations. 


\bibliographystyle{ACM-Reference-Format}
\bibliography{sample-base}


\end{document}